\documentclass[twocolumn,showpacs]{revtex4} %
\usepackage{epsfig,amsmath,amssymb,graphicx}
\usepackage[english]{babel}

\begin{document}

\title{ A SIMPLE MODEL OF BOSE-EINSTEIN CONDENSATION \\ OF INTERACTING PARTICLES} %
\author{Yu.M.\,Poluektov}
\email{yuripoluektov@kipt.kharkov.ua} %
\affiliation{%
National Science Center ``Kharkov Institute of Physics and
Technology'', 1, Akademicheskaya St., 61108 Kharkov, Ukraine }%

\begin{abstract}
A simple model of Bose-Einstein condensation of interacting
particles is proposed. It is shown that in the condensate state the
dependence of thermodynamic quantities on the interaction constant %
does not allow an expansion in powers of the coupling constant.
Therefore it is impossible to pass to the Einstein model of
condensation in an ideal Bose gas by means of a limiting passage,
setting the interaction constant to zero. The account for the
interaction between particles eliminates difficulties in the
description of condensation available in the model of an
ideal gas, which are connected with fulfilment of thermodynamic relations
and an infinite value of the particle number fluctuation
in the condensate phase. %
\newline%
\end{abstract}
\maketitle 

\section{Introduction}\vspace{-2mm}
Generalizing Bose's work \cite{Bose}, devoted to the statistics of
photons, to the case of particles with a finite mass, Einstein
\cite{Einstein} introduced a concept about condensation of particles
of an ideal gas in the momentum space. This effect, called
Bose-Einstein condensation, was subsequently used by F.\,London
\cite{London} and Tisza \cite{Tisza} to explain the phenomenon of
superfluidity of liquid helium discovered by Kapitsa \cite{Kapitsa}
and Allen \cite{Allen}. A new splash of interest to the phenomenon
of superfluidity was connected with the discovery about twenty years
ago of superfluidity in atomic gases of alkali metals confined in
magnetic \cite{PS,PS2} and laser traps \cite{Pit1}.

Although the phenomenon of Bose-Einstein condensation has been
studied for a long time both theoretically and experimentally, 
it cannot be considered entirely clear at present. Two main aspects
can be highlighted in this phenomenon. First, the condensation is
accompanied by the accumulation of a macroscopic number of particles
in the ground state. Second, as it has become clear later, the phase
transition to the condensate state is accompanied by the breaking of
the phase symmetry manifesting itself in the appearance of the complex field, 
that in fact is the cause of the phenomenon of superfluidity. The
effect of the accumulation of a macroscopic number of particles in
the state with the lowest energy is described already within the
scope of the Bose-Einstein model of an ideal gas \cite{Einstein}.
But, concerning the breaking of the phase symmetry, the theory of an
ideal Bose gas with the condensate does not provide its description,
because for that it is necessary to account for the interaction
between particles, which for the first time was done in the model of
a weakly nonideal Bose gas by Bogolyubov \cite{Bogolyubov}. In the
Bogolyubov approach the replacement of the operators of creation and
annihilation by a number $a_0^+\rightarrow \sqrt{N_0}$, $a_0\rightarrow \sqrt{N_0}$ %
leads to the breaking of the phase symmetry of Hamiltonian \cite{Bogolyubov}. %
Spatially inhomogeneous states of the Bose systems with broken phase
symmetry are well described by the Gross-Pitaevskii equation for the
complex function \cite{Gross,Pit2}.

Phase transitions in many-particle systems are a collective effect
and conditioned by the interaction between particles, therefore
phase transitions are absent in the model of an ideal gas. The only
exception, as it would seem, is Bose-Einstein condensation
which can be described by the model not accounting for 
the interparticle interaction. But the neglect of the interaction
leads to quite a number of considerable difficulties. Besides the
fact that the model of Bose-Einstein condensation of an ideal gas
does not account for the  breaking of the phase symmetry, the model
itself has a number of weak points. Thus, according to general
principles of statistical physics, the macroscopic properties of an
arbitrary physical system can be described with the help of one of
the thermodynamic potentials. It turns out often that the most
convenient is the use of the grand thermodynamic potential, whose
natural variables are the temperature and chemical potential. For
the condensate phase in the Einstein model we are forced to
postulate that the chemical potential equals zero, so that it ceases
to be an independent variable. Thus, a situation arises in which the
same system is described differently in different regions of the
phase diagram, with the help of different independent variables.
This seems unnatural and does not follow directly from the
postulates of thermodynamics, which is true regardless of the
microscopic nature of a considered system. Besides that, below the
condensation temperature the pressure proves to be a function of
only temperature and does not depend on the density. As a
consequence of this, the isobaric heat capacity \cite{P1} and the
isothermal compressibility become infinite \cite{LL}. A direct
indication of the limitation of the model of Bose-Einstein
condensation in an ideal gas is that the particle number fluctuation
in the condensate phase becomes infinite \cite{LL}.

In this paper, a model of Bose-Einstein condensation is proposed in
which the interaction between particles is taken into account. This
model, similarly to the Einstein model, describes only the effect of
the accumulation of particles in the ground state, but it does not
describe the phenomenon of the breaking of the phase symmetry,
because accounting for the breaking of the phase symmetry leads to a
considerable complication of the model \cite{P2,P3}. But even in the
proposed relatively simple model of condensation the account for
the interaction enables to eliminate the weaknesses of the model of
condensation in an ideal gas, which are indicated above.

In the proposed model the interparticle interaction is accounted for
in the self-consistent field approximation, and the phase transition
is described in a similar way as in the Einstein model. When
accounting for the interaction, both above and below the
condensation temperature, the system is characterized by the grand
thermodynamic potential for which the chemical potential is an
independent thermodynamic variable. The model does not have
difficulties connected with fulfilment of thermodynamic relations,
which are present in the model of an ideal Bose gas, and the
particle number fluctuation proves to be finite.

\begin{figure}[b!]
\vspace{0mm} \hspace{0mm}
\includegraphics[width = 0.97\columnwidth]{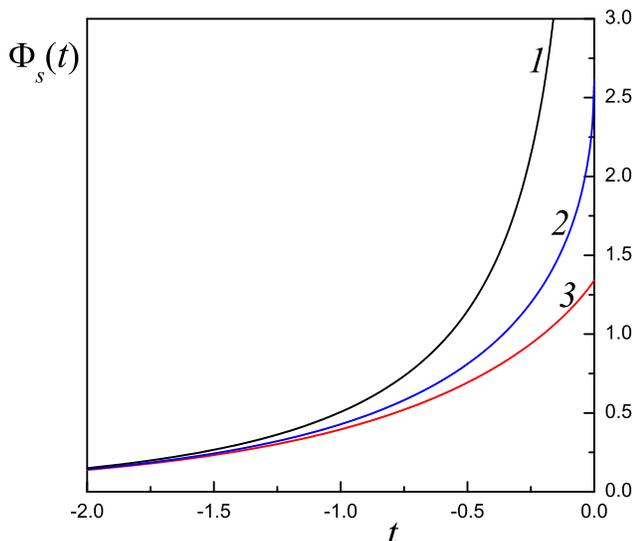} %
\vspace{-5mm}
\caption{\label{fig01} %
The function $\Phi_s(t)$ (\ref{01}) for different indexes: \newline %
1) $s=1/2$, 2) $s=3/2$, 3) $s=5/2$.
}%
\end{figure}

Before proceeding to the formulation of the model, we note that all
the thermodynamic characteristics of a Bose gas can be expressed
in terms of the special functions
\begin{equation} \label{01}
\begin{array}{l}
\displaystyle{%
  \Phi_s(t)=\frac{1}{\Gamma(s)}\int_0^{\!\infty} \frac{z^{s-1}\,dz}{e^{z-t}-1}, %
}%
\end{array}
\end{equation}
where $\Gamma(s)$ is the gamma function. The functions (\ref{01})
are defined for $t\leq 0$ and can be presented in the form of the
series $\Phi_s(t)=\sum_{n=1}^\infty e^{nt}/n^s$. For $t>0$, the
integral in (\ref{01}) diverges. For description of thermodynamic
properties of a Bose gas, the functions $\Phi_s(t)$ at
$s=1/2,\,3/2,\,5/2$ are generally sufficient. Note that at $s>1$, a
useful relation $\Phi_s'(t)=\Phi_{s-1}(t)$ is fulfilled. Besides,
$\Phi_{3/2}(0)=\zeta(3/2)=2.612$, $\Phi_{5/2}(0)=\zeta(5/2)=1.341$, %
and for $t\rightarrow -0\,$ $\Phi_{1/2}(t)\approx\sqrt{-\pi/t}$, %
where $\zeta(s)$ is the Riemann zeta function. The functions
$\Phi_s(t)$ steadily increase with increasing $t$, and for
$t\rightarrow -\infty$ the asymptotic $\Phi_s(t)\approx e^t$ holds
for them. The form of some of the functions (\ref{01}) is shown in
Fig.\,1.

\section{System of interacting Bose particles without the condensate}
In the self-consistent field model in the absence of the condensate
the many-particle system of bosons of mass $m$ is described by the
equation for the quasiparticle wave functions \cite{P2,P3}
\begin{equation} \label{02}
\begin{array}{l}
\displaystyle{%
  -\frac{\hbar^2}{2m}\Delta\phi_j({\bf r})+ \int\! d{\bf r}' W({\bf r},{\bf r}')\phi_j({\bf r}')=\varepsilon_j \phi_j({\bf r}), %
}%
\end{array}
\end{equation}
where the self-consistent field acting on a single particle is
determined by the expression
\begin{equation} \label{03}
\begin{array}{ll}
\displaystyle{%
  W\!({\bf r},{\bf r}')=\delta({\bf r}-{\bf r}')\int\!\! d{\bf r}''U({\bf r}-{\bf r}'')\rho({\bf r}'',{\bf r}'')\,+ %
}\vspace{2mm}\\ %
\displaystyle{%
\hspace{14mm}
  +\,U({\bf r}-{\bf r}')\rho({\bf r},{\bf r}'), %
} %
\end{array}
\end{equation}
where $\rho({\bf r},{\bf r}')=\sum_j \phi_j({\bf r})\phi_j^*({\bf r}')f_j$ is %
the one-particle density matrix, and
$f_j=\big[\!\exp(\beta\varepsilon_j) - 1\big]^{-1}$ is the
distribution function. The first term in (\ref{03}) describes the
direct interaction and the second one describes the exchange
interaction conditioned by the symmetry of the many-particle wave
function. In the spatially uniform case, which will be considered,
the equation (\ref{02}) is satisfied by solutions in the form of plane waves: %
$\phi_j({\bf r})\equiv\phi_{{\bf k}}({\bf r})=V^{-1/2}e^{i{\bf k}{\bf r}}$, %
$V$ is the system's volume, ${\bf k}$ is the wave vector, so that $j\equiv\{{\bf k}\}$. %
In case that particles interact through the delta-like potential
$U({\bf r}-{\bf r}')=g\delta({\bf r}-{\bf r}')$, the equation
(\ref{02}) leads to the following dispersion law of quasiparticles
\begin{equation} \label{04}
\begin{array}{ll}
\displaystyle{%
  \varepsilon_k = \frac{\hbar^2k^2}{2m}-\mu_*, %
}
\end{array}
\end{equation}
where, instead of the chemical potential $\mu$, the effective
dependent on the density chemical potential is present
\begin{equation} \label{05}
\begin{array}{ll}
\displaystyle{%
  \mu_* = \mu - 2\,g\,n, %
}
\end{array}
\end{equation}
where $n=N/V$ is the total particle number density. The condition
$t<0$ is equivalent to the condition for the chemical potential $\mu < 2\,g\,n$. %
As mentioned, the functions (\ref{01}) are defined for $t\leq 0$.
The cases $t<0$ and $t=0$ should be considered separately. Formally,
this is connected with the fact that the function $\Phi_{1/2}(t)$
tends to infinity for $t\rightarrow -0$ (Fig.\,1). In this section
the first possibility $t<0$ is considered, when the grand
thermodynamic potential as a function of the chemical potential and
temperature has the form
\begin{equation} \label{06}
\begin{array}{ll}
\displaystyle{%
  \Omega\equiv\Omega(\mu,T) = -V\!\left[ gn^2 + \frac{T}{\Lambda^3}\Phi_{5/2}(t) \right], %
}
\end{array}
\end{equation}
where $t\equiv\beta\mu_*=\beta(\mu-2gn)$, and the particle number
density
\begin{equation} \label{07}
\begin{array}{ll}
\displaystyle{%
  n=\frac{1}{V}\sum_k f_k = \frac{1}{\Lambda^3}\Phi_{3/2}(t). %
}
\end{array}
\end{equation}
Here $\Lambda\equiv\Lambda(T)\equiv\big(2\pi\hbar^2/mT\big)^{1/2}$ is the
de Broglie thermal wavelength. The formula (\ref{07}) defines the
particle number density as a function of the chemical potential and
temperature. It is easy to verify that the formula for the particle
number density (\ref{07}) follows from the thermodynamic relation
$N=-\big(\partial\Omega/\partial\mu\big)_T$. The entropy also can be
found both from the thermodynamic relation
$S=-\big(\partial\Omega/\partial T\big)_\mu$ and from the
combinatoric formula through the distribution function \cite{LL}:
\begin{equation} \label{08}
\begin{array}{ll}
\displaystyle{%
  S=\frac{V}{\Lambda^3}\!\left[ \frac{5}{2}\,\Phi_{5/2}(t) - t\,\Phi_{3/2}(t) \right]. %
}
\end{array}
\end{equation}

The formula for the pressure follows from (\ref{06}), taking into
account that $p=-\Omega/V$: %
\begin{equation} \label{09}
\begin{array}{ll}
\displaystyle{%
  p=gn^2 + \frac{T}{\Lambda^3}\,\Phi_{5/2}(t). %
}
\end{array}
\end{equation}
The formula for the energy can be obtained from the thermodynamic
relation $E=\Omega+TS+\mu N$:
\begin{equation} \label{10}
\begin{array}{ll}
\displaystyle{%
  E = V\!\left[ gn^2 + \frac{3}{2}\frac{T}{\Lambda^3}\Phi_{5/2}(t) \right]. %
}
\end{array}
\end{equation}
If we could, with the help of the formula (\ref{07}), exclude the
parameter $t$ from the formulas (\ref{08}) and (\ref{09}), we would
obtain the expressions for the entropy as a function of the volume,
the particle number and temperature and for the pressure as a
function of the density and temperature. But this is possible only
at high temperatures. In the general case the formulas (\ref{08})
and (\ref{09}), with account of (\ref{07}), define the entropy and
the pressure parametrically as functions of thermodynamic variables,
where the parameter $t$ varies in the range $-\infty<t<0$. The
formula for the entropy coincides with the expression for the case
of an ideal gas, and in the formula for the pressure a term with the
interaction constant appears. We note once more that the interaction
constant and the density enter into the definition of the parameter
$t=\beta(\mu-2gn)$.

Important directly measurable quantities are the heat capacities.
For their calculation in the general form it is convenient to use
the expressions for the total differentials of the entropy, the
particle number and the pressure:
\begin{equation} \label{11}
\begin{array}{ll}
\displaystyle{%
  dS=S\frac{dV}{V}+\frac{3}{2}S\frac{dT}{T}+\frac{V}{\Lambda^3}\!\left( \frac{3}{2}\,\Phi_{3/2} - t\,\Phi_{1/2} \right)\!dt, %
}
\end{array}
\end{equation}
\begin{equation} \label{12}
\begin{array}{ll}
\displaystyle{%
  dN=N\frac{dV}{V}+\frac{3}{2}N\frac{dT}{T}+\frac{V}{\Lambda^3}\Phi_{1/2}\,dt, %
}
\end{array}
\end{equation}
\begin{equation} \label{13}
\begin{array}{ll}
\hspace{-0mm}
\displaystyle{%
  dp=\!\frac{5}{2}\frac{\Phi_{5/2}}{\Lambda^3}dT\!+\!\frac{T}{\Lambda^3}\Phi_{3/2}dt+\!2gn^2\frac{dN}{N}\!-\!2gn^2\frac{dV}{V}. %
}
\end{array}
\end{equation}
Note that $d\Lambda/\Lambda=-dT/2T$. When calculating the heat
capacities, it should be kept in mind that the system with a fixed
number of particles is considered, so that $dN=0$. We should also
consider that $dV=0$ for the heat capacity at a constant volume
$C_V$, and $dp=0$ for the heat capacity at a constant pressure
$C_p$. As a result we find the heat capacity at a constant volume:
\begin{equation} \label{14}
\begin{array}{ll}
\displaystyle{%
  C_V=\frac{15}{4}\frac{V}{\Lambda^3}\!\left[ \Phi_{5/2}(t) - \frac{3}{5}\frac{\Phi_{3/2}^2(t)}{\Phi_{1/2}(t)} \right]. %
}
\end{array}
\end{equation}
The heat capacity at a constant pressure is determined by the formula:
\begin{equation} \label{15}
\begin{array}{ll}
\displaystyle{%
  C_p=\frac{25}{4}\frac{V}{\Lambda^3}\frac{\Phi_{5/2}(t)\Phi_{1/2}(t)}{\Phi_{3/2}^2(t)}\times
}\vspace{1mm}\\ %
\displaystyle{\hspace{08mm} %
  \times\!\left[ \Phi_{5/2}(t) - \frac{3}{5}\frac{\Phi_{3/2}^2(t)}{\Phi_{1/2}(t)} \right]\! %
  \frac{\left(\displaystyle{1+\frac{6}{5}\,\xi\frac{\Phi_{3/2}(t)}{\Phi_{5/2}(t)}}\right)}{\left(\displaystyle{1+2\,\xi\frac{\Phi_{1/2}(t)}{\Phi_{3/2}(t)}}\right)}. %
} %
\end{array}
\end{equation}
The interaction constant enters into the formula (\ref{15}) through
the parameter
\begin{equation} \label{16}
\begin{array}{ll}
\displaystyle{%
  \xi\equiv\xi(T,n)=\frac{g\,n}{T}. %
}
\end{array}
\end{equation}
The formula for the heat capacity $C_V$ (\ref{14}) coincides with
the respective formula for an ideal gas, and the expression for
$C_p$ contains explicitly the interaction constant, so that
(\ref{15}) turns into the formula for an ideal gas only for $g=0$.
In the limit of strong interaction $\xi\rightarrow\infty$ the heat
capacities coincide: $C_p\rightarrow C_V$. The difference of the
heat capacities is given by the formula: %
\begin{equation} \label{17}
\begin{array}{ll}
\displaystyle{\hspace{-0mm}%
  C_p-C_V=\frac{25}{4}\frac{V}{\Lambda^3}\frac{\Phi_{3/2}^2(t)}{\Phi_{1/2}(t)}\!
  \frac{\left[\displaystyle{\frac{\Phi_{5/2}(t)\Phi_{1/2}(t)}{\Phi_{3/2}^2(t)}-\frac{3}{5}}\right]^{\!2}}{\left(\displaystyle{1+2\,\xi\frac{\Phi_{1/2}(t)}{\Phi_{3/2}(t)}}\right)}. %
}
\end{array}
\end{equation}
It is easy to verify directly that the known thermodynamic identity
is satisfied:
\begin{equation} \label{18}
\begin{array}{ll}
\displaystyle{\hspace{-1mm}%
  C_p-C_V= -T\frac{\big(\partial p/\partial T\big)_V^2}{\big(\partial p/\partial V\big)_T}. %
}
\end{array}
\end{equation}

The calculation of the isothermal $\gamma_T$ and the adiabatic
$\gamma_\sigma$ compressibilities gives:
\begin{equation} \label{19}
\begin{array}{ll}
\displaystyle{\hspace{-1mm}%
  \gamma_T\equiv\frac{1}{n}\left(\frac{\partial n}{\partial p}\right)_T= %
  \frac{1}{nT}\left(\frac{\Phi_{3/2}(t)}{\Phi_{1/2}(t)}+2\,\xi\right)^{\!-1}, %
}\vspace{2mm}\\ %
\displaystyle{\hspace{-1mm}%
  \gamma_\sigma\equiv\frac{1}{n}\left(\frac{\partial n}{\partial p}\right)_\sigma= %
  \frac{3}{5nT}\left(\frac{\Phi_{5/2}(t)}{\Phi_{3/2}(t)}+\frac{6}{5}\,\xi\right)^{\!-1}. %
}
\end{array}
\end{equation}
Note that the entropy per one particle $\sigma=S/N$ depends only on
the parameter $t$, so that for the adiabatic processes $\sigma =
\textrm{const}$ also $t = \textrm{const}$. The square of speed of
sound is associated with the adiabatic compressibility by the
relation $u_\sigma^2=1/mn\gamma_\sigma$. The ratio of the
compressibilities (\ref{19}) coincides with the ratio of the heat
capacities:
\begin{equation} \label{20}
\begin{array}{ll}
\displaystyle{\hspace{-0mm}%
  \frac{\gamma_T}{\gamma_\sigma}=\frac{C_p}{C_V}=\frac{5}{3}
  \frac{\displaystyle{\left(\frac{\Phi_{5/2}(t)}{\Phi_{3/2}(t)}+\frac{6}{5}\,\xi\right)}}{\displaystyle{\left(\frac{\Phi_{3/2}(t)}{\Phi_{1/2}(t)}+2\,\xi\right)}}. %
}
\end{array}
\end{equation}

The derived formulas are simplified considerably at high
temperatures, for which $n\Lambda^3\ll 1$. Since in this limit
$\Phi_s(t)\approx e^t$, we have:
\begin{equation} \label{21}
\begin{array}{ll}
\displaystyle{\hspace{0mm}%
  n=\frac{e^t}{\Lambda^3},\qquad p=nT+gn^2, %
}\vspace{2mm}\\ %
\displaystyle{\hspace{0mm}%
  E=N\bigg(gn+\frac{3}{2}NT\bigg),\qquad S=N\ln\frac{e^{5/2}}{n\Lambda^3}, %
}\vspace{2mm}\\ %
\displaystyle{\hspace{0mm}%
  C_V=\frac{3}{2}N, \qquad C_p=\frac{5}{2}N\frac{\displaystyle{\left(1+\frac{6}{5}\,\xi\right)}}{\displaystyle{\left(1+2\,\xi\right)}}, %
}\vspace{2mm}\\ %
\displaystyle{\hspace{0mm}%
  \gamma_T=\frac{1}{nT}\left(1+2\,\xi\right)^{-1},
}\vspace{2mm}\\ %
\displaystyle{\hspace{0mm}%
  \gamma_\sigma=\frac{1}{mnu_\sigma^2}=\frac{3}{5nT}\!\left(1+\frac{6}{5}\,\xi\right)^{\!-1}. %
}
\end{array}
\end{equation}
With neglect of the interaction these relations, naturally, turn
into the formulas for a classical ideal monoatomic gas.

\section{System of interacting Bose particles with the condensate}
Here we consider the case $t=0$ separately. In this case, according
to (\ref{05}), the chemical potential and the total particle number
density are connected by the relation
\begin{equation} \label{22}
\begin{array}{ll}
\displaystyle{%
  \mu = 2\,g\,n. %
}
\end{array}
\end{equation}
Note that this formula differs by a factor of 2 from the respective
formulas in the Gross-Pitaevskii theory \cite{Gross,Pit2} or in the
approaches based on the replacement of the operators of creation and
annihilation of particles in the condensate by a $c$\,-\,number
\cite{Beliaev,HP}. This is connected with the fact that, as seen
from the relation (\ref{03}), both the direct and exchange
interactions are accounted for here giving the same contribution for
the point interaction, whereas the exchange interaction is not 
accounted for in the mentioned approaches \cite{Gross,Pit2,Beliaev,HP}. %

In contrast to the case of an ideal Bose gas where it is assumed
that $\mu=0$ in the condensate state, in this case the effective
chemical potential (\ref{05}) becomes zero and the real chemical
potential remains to be a correct independent variable. Therefore
the system, the same as in the absence of the condensate, can be
characterized by the grand thermodynamic potential expressed in the
variables of the chemical potential $\mu$ and temperature $T$:
\begin{equation} \label{23}
\begin{array}{ll}
\displaystyle{%
  \Omega\equiv\Omega(\mu,T) = -V\!\left[ \frac{\mu^2}{4g} + \frac{T}{\Lambda^3}\zeta(5/2)\right]. %
}
\end{array}
\end{equation}
Naturally, in this case as well the total particle number and the
entropy are determined by the usual thermodynamic formulas
$N=-\big(\partial\Omega/\partial\mu\big)_T$ and
$S=-\big(\partial\Omega/\partial T\big)_\mu$. The energy of the
condensate phase is given by the formula:
\begin{equation} \label{24}
\begin{array}{ll}
\displaystyle{%
  E = V\!\left[ gn^2 + \frac{3}{2}\frac{T}{\Lambda^3}\zeta(5/2) \right]. %
}
\end{array}
\end{equation}
Pay attention that the interaction constant enters into the
denominator of the thermodynamic potential (\ref{23}), which is, as
will be discussed below, a very essential fact. And the particle
number density, calculated with the help of the distribution
function by the formula
\begin{equation} \label{25}
\begin{array}{ll}
\displaystyle{%
  n' = \frac{1}{\Lambda^3}\zeta(3/2), %
}
\end{array}
\end{equation}
depends on temperature and decreases with decreasing temperature. In
the case of a system with a fixed density, following Einstein's idea
\cite{Einstein}, we have to assume that the total density is a sum
of the overcondensate particle number density (\ref{25}) and the
density of particles in the state with the lowest energy $n_0$, so
that $n=n'+n_0$. %
Temperature $T_B$, at which the particle number density determined by
the formula (\ref{25}) coincides with the total density, is the
critical temperature of Bose-Einstein condensation:
\begin{equation} \label{26}
\begin{array}{ll}
\displaystyle{%
   T_B=\frac{2\pi\hbar^2}{m}\!\left[ \frac{n}{\zeta(3/2)} \right]^{\!2/3}. %
}
\end{array}
\end{equation}
At this temperature the density and the de Broglie wavelength are
connected by the relation $n\Lambda_B^3=\zeta(3/2)$. In the case of
the point interaction the critical temperature (\ref{26}) coincides
with the condensation temperature in an ideal gas. %
For the nonlocal interparticle interaction potential the formula
(\ref{26}) will contain the effective mass, but in this paper we
confine ourselves to consideration of the point interaction. Thus,
the particle number density in the condensate as a function of
temperature is determined by the same formula as in an ideal gas:
\begin{equation} \label{27}
\begin{array}{ll}
\displaystyle{%
   n_0(T)=n\!\left[ 1- \left(\frac{T}{T_B}\right)^{\!3/2} \right]. %
}
\end{array}
\end{equation}

The entropy and the pressure below the transition temperature are
determined by the formulas:
\begin{equation} \label{28}
\begin{array}{ll}
\displaystyle{%
  S = \frac{5}{2}\frac{V}{\Lambda^3}\zeta(5/2), %
}
\end{array}
\end{equation}
\begin{equation} \label{29}
\begin{array}{ll}
\displaystyle{%
  p = gn^2 + \frac{T}{\Lambda^3}\zeta(5/2). %
}
\end{array}
\end{equation}
Here the formula for the entropy coincides with the case of an ideal
gas, but the pressure depends not only on temperature as in an ideal
gas but on the density as well. The contribution in the pressure
from the density proves to be twice larger than that in the
Gross-Pitaevskii theory, which is conditioned as remarked above by
accounting for the exchange interaction in the self-consistent field
(\ref{03}). The entropy per one overcondensate particle
below the transition temperature ($N'=(V/\Lambda^3)\zeta(3/2)$)
does not depend on the thermodynamic variables and the mass of particles,
being a universal constant (in Boltzmann's constant units)
for which we introduce a special designation:
\begin{equation} \label{30}
\begin{array}{ll}
\displaystyle{%
  \sigma_0=\frac{S}{N'}= \frac{5}{2}\frac{\zeta(5/2)}{\zeta(3/2)}\approx 1.283. %
}
\end{array}
\end{equation}

The heat capacity at a constant volume is the same as in the case of an ideal gas
\begin{equation} \label{31}
\begin{array}{ll}
\displaystyle{%
  C_V= \frac{3}{2}N\sigma_0\!\left(\frac{T}{T_B}\right)^{\!3/2}, %
}
\end{array}
\end{equation}
and the isobaric heat capacity has the form
\begin{equation} \label{32}
\begin{array}{ll}
\displaystyle{%
  C_p= \frac{3}{2}N\sigma_0\!\left(\frac{T}{T_B}\right)^{\!3/2}\!\left[ 1+\frac{\sigma_0}{3\xi_B}\!\left(\frac{T}{T_B}\right)^{\!5/2} \right], %
}
\end{array}
\end{equation}
where $\xi_B\equiv\xi(T_B,n)=gn/T_B$ is the parameter (\ref{16})
at the condensation temperature. The difference of the heat capacities
is given by the formula
\begin{equation} \label{33}
\begin{array}{ll}
\displaystyle{%
  C_p-C_V = \frac{N\sigma_0^2}{2\xi_B}\!\left(\frac{T}{T_B}\right)^{\!4}. %
}
\end{array}
\end{equation}
These temperature dependencies are in accordance with
the thermodynamic requirements for the behavior of heat capacities at $T\rightarrow 0$,
namely for $S\sim T^n$ it should be $C_p-C_V \sim T^{2n+1}$ and
$(C_p-C_V)/C_p \sim T^{n+1}$ \cite{LL}. In the present case $n=3/2$.

It is easy to make sure that the thermodynamic identity (\ref{18})
is satisfied in the condensate phase as well. In an ideal Bose gas
at $T<T_B$ the pressure does not depend on the volume, so that
the denominator of the right part of this formula becomes zero.
Since for $g\rightarrow 0$ the isobaric heat capacity $C_p$
becomes infinite \cite{P1}, then both the left and right parts of the identity (\ref{18})
become infinite for an ideal gas. The condition of thermodynamic stability
requires that the difference of the heat capacities (\ref{33}) be positive,
so that the performed consideration is valid only for the case $g>0$,
that is when the interparticle interaction is primarily of a repulsive character.
Note that the fulfilment of this condition is required both in the Bogolyubov theory of
a weakly nonideal Bose gas \cite{Bogolyubov} and in the Gross-Pitaevskii theory \cite{Gross,Pit2}.

The isothermal and adiabatic compessibilities at $T<T_B$ are determined by the formulas: %
\begin{equation} \label{34}
\begin{array}{ll}
\displaystyle{\hspace{0mm}%
  \gamma_T=\frac{1}{2gn^2},\quad
  \gamma_\sigma=\frac{1}{2gn^2}\!\left[ 1+\frac{\sigma_0}{3\xi_B}\!\left(\frac{T}{T_B}\right)^{\!5/2} \right]^{\!-1}. %
}
\end{array}
\end{equation}
In the condensate phase the ratio of the compressibilities
also coincides with the ratio of the heat capacities:
\begin{equation} \label{35}
\begin{array}{ll}
\displaystyle{\hspace{-0mm}%
  \frac{\gamma_T}{\gamma_\sigma}=\frac{C_p}{C_V}= 1+\frac{\sigma_0}{3\xi_B}\!\left(\frac{T}{T_B}\right)^{\!5/2}. %
}
\end{array}
\end{equation}
The speed of sound $u_\sigma$ in the condensate phase has the form
\begin{equation} \label{36}
\begin{array}{ll}
\displaystyle{\hspace{0mm}%
  u_\sigma^2 =\frac{2gn}{m}\!\left[ 1+\frac{\sigma_0}{3\xi_B}\!\left(\frac{T}{T_B}\right)^{\!5/2} \right]. %
}
\end{array}
\end{equation}
The temperature-dependent contribution in the speed of sound is
determined by the overcondensate particles. With neglect of the
interparticle interaction, the speed of sound tends to zero as
$T\rightarrow 0$. The speed of sound due to the interaction,
to within a factor of 2 that appears as mentioned due to
accounting for the exchange interaction, coincides with
the expression derived by Bogolyubov \cite{Bogolyubov}.

It is convenient to express the interaction constant $g$ in terms of
a directly observable quantity -- the scattering length $a$:
\begin{equation} \label{37}
\begin{array}{ll}
\displaystyle{\hspace{0mm}%
 g =\frac{4\pi\hbar^2a}{m}. %
}
\end{array}
\end{equation}
Thus, three characteristic lengths can be distinguished
in the gas of interacting particles: the temperature-dependent
de Broglie thermal wavelength $\Lambda$, the average distance
between particles $l=n^{-1\!/3}$ and the scattering length $a$.
The ratio $q\equiv\Lambda/l$ characterizes the role of quantum effects,
that is determines the degree of ``quantumness'' of the system.
The larger this ratio, the more considerable is the role of quantum effects.
At the transition temperature this ratio equals
$q_B\equiv\Lambda(T_B)/l=[\zeta(3/2)]^{1\!/3}\approx 1.38$,
that is the de Broglie wavelength somewhat exceeds the average distance
between particles. The parameter (\ref{16}) can be written in the form
$\xi=\xi_B(T_B/T)$, where the dimensionless quantity
determining the role of the interparticle interaction at the temperature
of transition into the condensate state:
\begin{equation} \label{38}
\begin{array}{ll}
\displaystyle{\hspace{0mm}%
 \xi_B\equiv 2\frac{\Lambda_B^2a}{l^3}. %
}
\end{array}
\end{equation}

As follows from the formulas (\ref{32}) and (\ref{34}),
the isobaric heat capacity and the isothermal compressibility
prove to be finite only when the interaction between particles
is taken into account. In the model of an ideal gas both of
these quantities become infinite \cite{P1}. This is a trivial
consequence of the fact that in an ideal gas with the condensate
the pressure depends only on temperature but not on the density.
Therefore, the adding of heat to an ideal gas with the condensate
at a constant pressure causes not the increase of temperature
but the ``evaporation'' of the condensate \cite{P1}. The account for
the interaction between particles leads to finite values of
the isobaric heat capacity and the isothermal compressibility
and to fulfilment of all the thermodynamic relations
in the phase with Bose-Einstein condensate.

\section{Bose system near the transition temperature}
The most interesting is naturally the behavior of thermodynamic
quantities near the temperature of transition into the condensate state.
Here in calculations one should make use of the expansions
\begin{equation} \label{39}
\begin{array}{cc}
\displaystyle{\hspace{0mm}%
\Phi_{1/2}(t)\approx\!\sqrt{-\frac{\pi}{t}}+\zeta(1/2),\,\,\, \Phi_{3/2}(t)\approx\zeta(3/2)-2\sqrt{-\pi t},%
}\vspace{2mm}\\ %
\displaystyle{\hspace{0mm}%
\Phi_{5/2}(t)\approx\zeta(5/2)+\zeta(1/2)\,t,%
}
\end{array}
\end{equation}
being valid for $|t|\ll 1$. At $T\geq T_B$ the expansions in
the parameter $\tau=(T-T_B)/T_B$ of the entropy and the pressure
at a fixed number of particles and their density $n=N/V$ have the form
\begin{equation} \label{40}
\begin{array}{ll}
\displaystyle{%
  S = N\sigma_0\!\left[1+\frac{3}{2}\,\tau+\frac{3}{8}\!\left(1-\frac{9}{10}\,\alpha\right)\!\tau^2\right], %
}
\end{array}
\end{equation}
\vspace{-2mm}
\begin{equation} \label{41}
\begin{array}{ll}
\displaystyle{%
  p = gn^2 + p_0\!\left[1+\frac{5}{2}\,\tau+\frac{15}{8}\!\left(1-\frac{3}{10}\,\alpha\right)\!\tau^2\right], %
}
\end{array}
\end{equation}
where $p_0\equiv(2/5)\,\sigma_0nT_B$ is the pressure
at the transition temperature in an ideal Bose gas,
$\alpha\equiv[\zeta(3/2)]^3\big/\pi\zeta(5/2)\approx 4.230$. %

The behavior of the heat capacity at a constant volume
is the same as in the case of an ideal gas:
\begin{equation} \label{42}
\begin{array}{ll}
\displaystyle{%
  C_V= \frac{3}{2}N\sigma_0\!\!\left[1+\frac{3}{2}\!\left(1-\frac{3}{10}\,\alpha\right)\!\tau\right], %
}
\end{array}
\end{equation}
and the heat capacity at a constant pressure has the form
\begin{equation} \label{43}
\begin{array}{ll}
\displaystyle{%
  C_p= C_{pB}\!\big[1+B\tau\big], %
}
\end{array}
\end{equation}
where
\begin{equation} \label{44}
\begin{array}{ll}
\displaystyle{%
  C_{pB}= \frac{3}{2}\,N\sigma_0\!\left[1+\frac{\sigma_0}{3\xi_B}\right] %
}
\end{array}
\end{equation}
is the isobaric heat capacity at the transition temperature.
The coefficient in (\ref{43}) has the form
\begin{equation} \label{45}
\begin{array}{ll}
\displaystyle{\hspace{-0mm}%
 B= \frac{\displaystyle{\frac{3}{2}\!\left(1-\frac{3}{10}\,\alpha\right)\!+\!\frac{4}{3\xi_B}\!\left(1-\frac{9}{80}\,\alpha\right)}}{\displaystyle{1+\frac{\sigma_0}{3\xi_B}}} %
 -\frac{3\zeta^2(3/2)}{8\pi\xi_B}.
}
\end{array}
\end{equation}

\begin{figure}[b!]
\vspace{-2mm} \hspace{0mm}
\includegraphics[width = 0.97\columnwidth]{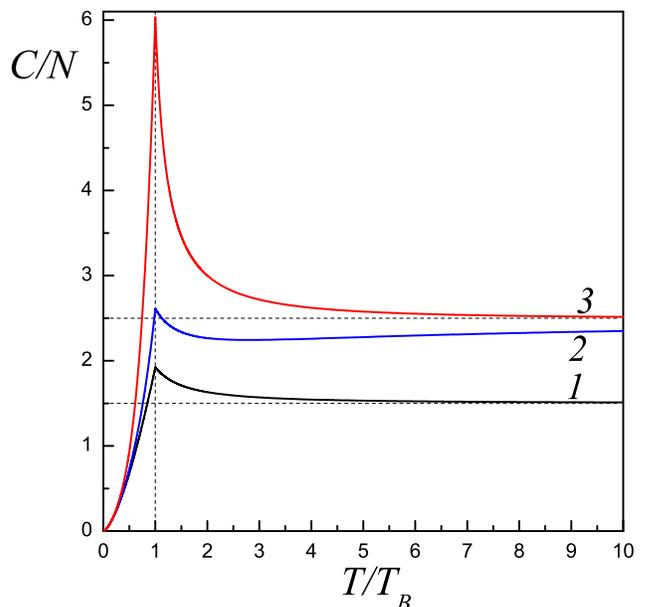} %
\vspace{-5mm}
\caption{\label{fig02} %
Dependencies of the heat capacities per one particle on temperature:
1) $C_V(T)$; 2) $C_p(T)$ for $\xi_B=1.2$; 3) $C_p(T)$ for $\xi_B=0.2$.%
}%
\end{figure}

In an ideal gas, with approaching to the condensation temperature
from the side of higher temperatures, the isobaric heat capacity
tends to infinity \cite{P1}:
\begin{equation} \label{46}
\begin{array}{ll}
\displaystyle{%
  C_p\approx\frac{10}{3}\frac{N\sigma_0}{\alpha\tau}. %
}
\end{array}
\end{equation}
The both heat capacities are continuous at the transition temperature,
but their derivatives at the transition from the high-temperature to the low-temperature phase
$\Delta\big(\partial C/\partial T\big)\equiv\big(\partial C/\partial T\big)_{T_B+0}-\big(\partial C/\partial T\big)_{T_B-0}$ %
undergo jumps:
\begin{equation} \label{47}
\begin{array}{ll}
\displaystyle{%
  \Delta\!\left(\frac{\partial C_V}{\partial T}\right)= -\frac{27}{16\pi}\zeta^2(3/2)\frac{N}{T_B}, %
}
\end{array}
\end{equation}
\begin{equation} \label{48}
\begin{array}{ll}
\displaystyle{%
  \Delta\!\left(\frac{\partial C_p}{\partial T}\right)= -\frac{27}{16\pi}\zeta^2(3/2)\frac{N}{T_B}\!\left[1+\frac{\sigma_0}{3\xi_B}\right]^{\!2}. %
}
\end{array}
\end{equation}
The jump of the derivative of the isobaric heat capacity increases with
decreasing the interaction strength, and in the limit of the strong interaction
$\xi_B\gg 1$ the jumps of the derivatives for the both heat capacities
(\ref{47}) and (\ref{48}) coincide. Some dependencies of the heat capacities
on temperature are presented in Fig.\,2. Attention should be paid that
for a rather large value of the interaction the dependence $C_p(T)$ at
$T>T_B$ has a minimum (curve 2, Fig.\,2), which corresponds qualitatively
to the observable analogous dependence in the liquid helium \cite{BF}. %
The isothermal compressibility is continuous at the condensation temperature,
and its derivative with respect to temperature undergoes a jump:
\begin{equation} \label{49}
\begin{array}{ll}
\displaystyle{%
  \Delta\!\left(\frac{\partial \gamma_T}{\partial T}\right)= -\frac{3\zeta^2(3/2)}{16\pi g^2n^3}. %
}
\end{array}
\end{equation}
The temperature dependencies of the isothermal compressibility and
the square of speed of sound are shown in Fig.\,3.
\begin{figure}[t!]
\vspace{2mm} \hspace{0mm}
\includegraphics[width = 0.99\columnwidth]{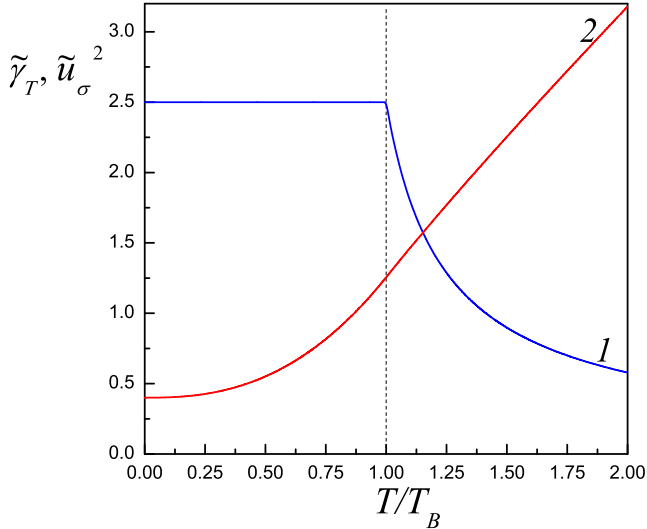} %
\vspace{-8.0mm}
\caption{\label{fig03} %
Temperature dependencies of: %
1) the isothermal compressibility $\tilde{\gamma}_T(T)\equiv\gamma_T(T)nT_B$ and %
2) the square of speed of sound $\tilde{u}_\sigma^2(T)=u_\sigma^2(T)(m/T_B)$ for the parameter $\xi_B=0.2$. %
}%
\end{figure}

Thus, in this model the transition into the condensate state,
the same as in the case of an ideal gas, is the phase transition of the third kind.
A somewhat different, more questionable as it seems to us, interpretation
of this transition as the phase transition of the first kind is given in the book \cite{Huang}.

\section{The particle number fluctuation}\vspace{-2mm}
As known \cite{LL}, below the temperature of Bose-Einstein condensation
the particle number fluctuation in an ideal Bose gas becomes infinite,
which directly indicates the necessity of accounting for the interparticle
interaction for a correct description of the condensate state.
The particle number fluctuation can be calculated by the known formula \cite{LL}: %
\begin{equation} \label{50}
\begin{array}{ll}
\displaystyle{%
  \big\langle(\Delta N)^2\big\rangle = T\!\left(\frac{\partial N}{\partial \mu}\right)_{\!T,V}. %
}
\end{array}
\end{equation}

Calculation above and below the transition temperature gives:
\begin{equation} \label{EQ51}
\begin{array}{ll}
\displaystyle{%
 \left(\frac{\partial N}{\partial \mu}\right)_{\!T,V}=
 \left\{
 \begin{array}{l}
   \frac{\displaystyle{N}}{\displaystyle{ T\frac{\Phi_{3/2}(t)}{\Phi_{1/2}(t)}+2gn } }, \quad T>T_B,  \vspace{1mm} \\ %
   \hspace{10mm}
   \frac{\displaystyle{N}}{\displaystyle{2gn}}, \hspace{12mm} T<T_B.
 \end{array}
 \right.
}%
\end{array}
\end{equation}
As seen, at the condensation temperature these formulas coincide,
so that the particle number fluctuation changes continuously
at the transition into the condensate state.
The relative particle number fluctuations
$\delta_N\equiv\sqrt{\big\langle(\Delta N)^2\big\rangle}\Big/N$
above and below the transition temperature are determined by the formulas:
\begin{equation} \label{EQ52}
\begin{array}{ll}
\displaystyle{%
 \delta_N=
 \left\{
 \begin{array}{l}
   \frac{\displaystyle{1}}{\displaystyle{\sqrt{N}}}\frac{\displaystyle{1}}{\displaystyle{\sqrt{2\xi+\frac{\Phi_{3/2}(t)}{\Phi_{1/2}(t)}} } }, \quad T>T_B,  \vspace{1.5mm} \\ %
   \hspace{0mm}
   \frac{\displaystyle{1}}{\displaystyle{\sqrt{N}}}\frac{\displaystyle{1}}{\displaystyle{\sqrt{2\xi_B}}}\left(\frac{\displaystyle{T}}{\displaystyle{T_B}}\right)^{\!1/2}, \hspace{3.0mm} T<T_B.
 \end{array}
 \right.
}%
\end{array}
\end{equation}
Dependencies of the relative particle number fluctuation
on temperature for the cases of strong and weak interaction are shown in Fig.\,4.
In the limit of high temperatures, such that $T\gg 2gn$ and $n\Lambda^3\ll\zeta(3/2)$, %
the relative fluctuation is the same as in an ideal gas $\delta_N=1\big/\sqrt{N}$. %
Near the transition temperature and in the condensate phase
the relative fluctuation is substantially determined by the value of interaction:
the weaker the interaction, the larger is the fluctuation near the transition temperature.
The derivative of the relative fluctuation on temperature undergoes a jump at $T_B$:
\begin{equation} \label{53}
\begin{array}{ll}
\displaystyle{%
  \Delta\!\left(\frac{\partial \delta_N}{\partial T}\right)= -\frac{3\zeta^2(3/2)}{16\pi\sqrt{2}}\frac{1}{\xi_B^{3\!/2}\sqrt{N}}. %
}
\end{array}
\end{equation}
Thus, as it should be expected, the account for the interparticle interaction
eliminates a serious drawback in the model of condensation in an ideal Bose gas
consisting in an infinite value of the fluctuation of the number of particles in the condensate phase.
\begin{figure}[h!]
\vspace{0mm} \hspace{-0mm}
\includegraphics[width = 0.97\columnwidth]{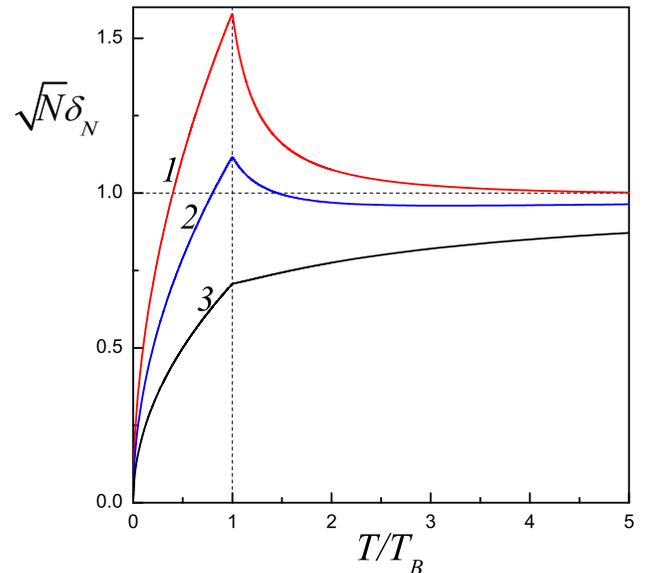} %
\vspace{-5.0mm}
\caption{\label{fig04} %
Temperature dependencies of the relative particle number fluctuation
$\delta_N\!\sqrt{N}$ for different values of the interaction: %
1) $\xi_B=0.2$; 2) $\xi_B=0.4$; 3) $\xi_B=1.0$.
}%
\end{figure}

\vspace{-5mm}
\section{Conclusion}\vspace{-2mm}
The model of Bose-Einstein condensation of interacting
particles, formulated in this paper, allows to
make some conclusions about the reason of actual nonapplicability
of the model of an ideal gas for the Bose systems at low temperatures.
A system of a large number of particles, interacting via the pair potential
$U({\bf r}-{\bf r}')=g\,u({\bf r}-{\bf r}')$, can be described
with the help of the grand thermodynamic potential $\Omega=\Omega(T,\mu;g)$ %
which is a function of the independent thermodynamic variables $T,\mu$
and the interaction constant $g$. The total number of particles
in the system $N=-\big(\partial\Omega/\partial\mu\big)_{V,T}=N(T,\mu;g)$
is also a function of these quantities. In order for the transition to be possible
to the model of an ideal gas in some range of values of the variables $T,\mu$
in the system with a constant volume and a fixed number of particles $N_0$,
the finite limits must exist 
\begin{equation} \label{54}
\begin{array}{ll}
\displaystyle{\hspace{-2mm}%
  \lim_{g\rightarrow 0}\Omega(T,\mu;g)=\Omega_0(T,\mu),\quad \lim_{g\rightarrow 0}N(T,\mu;g)=N_0. %
}
\end{array}
\end{equation}

A many-particle system of interacting particles can be in different phase states
for many of which the limiting passages (\ref{54}) can either not exist.
Thus, for example, in the case of a normal Fermi liquid such limiting passages
are possible. At that the limit $g\rightarrow 0$ means ``switching off''
the interaction and the transition from a Fermi liquid to a Fermi gas.
But if a Fermi system is in the superfluid or the superconducting state,
then transitions from these states into the state of noninteracting particles
are impossible. This is conditioned by that here
the dependence of the binding energy on the interaction constant is such
that it cannot be expanded in powers of $g$. In its time this
algebraic difficulty was a considerable obstacle for development
of the superconductivity theory \cite{Gennes}. 

As seen from the expression for the thermodynamic potential (\ref{23}),
completely the same situation takes place in the case of
the system of Bose particles in the presence of the condensate.
A feature of the Bose systems is that for them there do not exist phases
at low temperatures, from which the transition to the noninteracting system
could be possible by means of the limiting passages (\ref{54}).
Indeed, the interaction constant in the thermodynamic potential (\ref{23})
is in the denominator, so that for $g\rightarrow 0$ the potential becomes
infinite. Formally this difficulty can be bypassed,
that Einstein has done \cite{Einstein}, by fixing the chemical potential,
setting it zero and thus passing to consideration of
the system with a variable number of particles. Then the term
which is singular in the interaction constant drops out from
the thermodynamic potential (\ref{23}). However,
as discussed in this paper, this leads to difficulties connected
with the consistency of thermodynamic description and the fulfilment of
some thermodynamic relations, as well as to an infinite value of
the fluctuation of the number of particles. One more peculiarity of
the Bose systems consists in that the transition temperature into
the condensate state (\ref{26}), in contrast to the superfluid Fermi systems,
does not contain the interaction constant. This somewhat conceals
a fundamental importance of accounting for the interaction
between particles at the transition into the condensate phase.

In this paper it is shown on the example of a simple model how
the account for the interaction between particles enables to eliminate
the difficulties of the model of an ideal Bose gas. Although
the presence of the singularity in the interaction constant in
the thermodynamic potential (\ref{23}) is demonstrated in a simplified
model, this is in fact true in more realistic models.
Notice also that the performed consideration indicates that
the perturbation theory for the Bose systems with the condensate, which is
based on the choice of the model of an ideal gas as the main approximation,
cannot be consistent \cite{Beliaev,HP}. In constructing the perturbation theory
the effects of the interaction between particles should be approximately
taken into account already in the main approximation,
as it can be done, for example, in the self-consistent field model \cite{P3}.



\vspace{-0mm}
\begin{thebibliography}{99}
\bibitem{Bose}
  S.N.\,Bose, Plancks gesetz und lichtquanten hypothese, Z. Phys.
  \textbf{26}(1), 178\,--\,181 (1924).
\bibitem{Einstein}
  A.\,Einstein, Quantum theory of the monatomic ideal gas,
  Sitzungsberichte der Preussischen Akademie der Wissenschaften,
  Physikalisch-mathematische Klasse, 261\,--\,267, 1924; 3\,--\,14,
  1925. In a book: \newline A.\,Einstein, A collection of scientific
  works, Vol.\,3, Nauka, Moscow, 481\,--\,511 (1966).
\bibitem{London}
  F.\,London, The $\lambda$\,-\,phenomenon of liquid helium and the
  Bose-Einstein degeneracy, Nature \textbf{141}, 643 (1938).
\bibitem{Tisza}
  L.\,Tisza, Transport phenomena in helium II, Nature \textbf{141}, 913 (1938). %
\bibitem{Kapitsa}
  P.L.\,Kapitsa, Viscosity of liquid helium below the $\lambda$-point,
  Nature \textbf{141}, 74 (1938).
\bibitem{Allen}
  J.F.\,Allen, H.\,Jones, New phenomena connected with heat flow in
  helium II, Nature \textbf{141}, 234 (1938).
\bibitem{PS}
  L.\,Pitaevskii, S.\,Stringari, Bose-Einstein condensation, Oxford
  University Press, USA, 492 p. (2003).
\bibitem{PS2}
  C.H.\,Pethick, H.\,Smith,  Bose-Einstein condensation in dilute
  gases, Cambridge University Press, 402 p. (2002).
\bibitem{Pit1}
  L.P.\,Pitaevskii, Bose-Einstein condensates in a laser radiation
  field, Phys. Usp. \textbf{49}, 333\,--\,351 (2006).
\bibitem{Bogolyubov}
  N.N.\,Bogolyubov, On the theory of superfluidity, J. Phys. USSR
  \textbf{11}, 23\,--\,32 (1947); Izv.\,AN SSSR, Ser. Fiz. \textbf{11}, 77\,--\,90 (1947). 
\bibitem{Gross}
  E.P.\,Gross, Structure of a quantized vortex in boson systems, Il
  Nuovo Cimento \textbf{20}, 454\,--\,477 (1961).
\bibitem{Pit2}
  L.P.\,Pitaevskii, Vortex lines in an imperfect Bose gas, Sov. Phys.
  JETP \textbf{13}, 451\,--\,454 (1961).
\bibitem{P1}
  Yu.M.\,Poluektov, Isobaric heat capacity of an ideal Bose gas, Russ.
  Phys. J. \textbf{44}\,(6), 627\,--\,630 (2001).
\bibitem{LL}
  L.D.\,Landau, E.M.\,Lifshitz, Statistical physics, Vol. 5 (Part 1),
  Butterworth-Heinemann, Oxford, 544 p. (1980).
\bibitem{P2}
  Yu.M.\,Poluektov, Self-consistent field model for spatially
  inhomogeneous Bose systems, Low Temp. Phys. \textbf{28}, 429\,--\,441 (2002). 
\bibitem{P3}
  Yu.M.\,Poluektov, On the quantum-field description of many-particle Bose systems with spontaneously broken symmetry, %
  Ukr. J. Phys. \textbf{52}\,(6), 579\,--\,595 (2007); arXiv:1306.2103[cond-mat.stat-mech]. %
\bibitem{Beliaev}
  S.T.\,Beliaev, Application of the methods of quantum field theory to
  a system of bosons, Sov. Phys. JETP \textbf{7}, 289\,--\,299 (1958). 
\bibitem{HP}
  N.M.\,Hugenholtz, D.\,Pines, Ground-state energy and excitation
  spectrum of a system of interacting bosons, Phys. Rev. \textbf{116}, 489\,--\,506 (1959). %
\bibitem{BF}
  M.\,Buckingham, W.\,Fairbank, The nature of lambda-transition in
  liquid helium, Prog. in Low Temp. Phys. \textbf{3}, 80\,--\,112 (1961). 
\bibitem{Huang}
K.\,Huang, Statistical\,mechanics, Wiley, 493\,p. (1987). %
\bibitem{Gennes}
  P.G.\,de\,Gennes, Superconductivity of metals and alloys,
  Benjamin, New York, 274 p. (1966). 
\end{thebibliography}
\end{document}